\documentclass[12pt]{article}
\usepackage{epsf}
\usepackage[dvips]{color} 
\usepackage{exscale}
\usepackage[dvips]{graphicx}
\usepackage{amsmath}
\usepackage{amssymb}
\usepackage{latexsym}
\usepackage{cite}
\usepackage{color}

\begin{document}
\setlength{\baselineskip}{18pt}
\begin{titlepage}
\begin{flushright}
{\small KANAZAWA-15-08}\\
{\small SU-HET-06-2015}
\end{flushright}

\vspace*{1.2cm}

\begin{center}

{\Large\bf A model realizing inverse seesaw and resonant leptogenesis}
\lineskip .75em
\vskip 1.5cm

\normalsize
{\large Mayumi Aoki}$^1$,
{\large Naoyuki Haba}$^2$, and 
{\large Ryo Takahashi}$^{2,3}$

\vspace{1cm}

$^1${\it Institute for Theoretical Physics, Kanazawa University, 

Kanazawa, 920-1192, Japan}\\

$^2${\it Graduate School of Science and Engineering, Shimane University, 

Matsue, 690-8504, Japan}\\

$^3${\it Graduate School of Science, Tohoku University, 

Sendai, 980-8578, Japan}

\vspace*{10mm}

{\bf Abstract}\\[5mm]
{\parbox{13cm}{\hspace{5mm}
We construct a model realizing the inverse seesaw mechanism. The model has two 
types of gauge singlet fermions in addition to right-handed 
neutrinos. A required Majorana mass scale (keV scale) for generating the light 
active neutrino mass in the conventional 
inverse seesaw can be naturally explained
by a ``seesaw'' mechanism between the two singlet fermions in our 
model. We find that our model can decrease the magnitude of hierarchy among the mass
 parameters by $\mathcal{O}(10^4)$ from that in the conventional inverse seesaw 
model. We also show that a successful resonant leptogenesis occurs for 
generating the baryon asymmetry of the universe in our model. The desired mass 
degeneracy for the resonant leptogenesis can also be achieved by the ``seesaw''
between the two singlet fermions.
}}

\end{center}
\end{titlepage}

\section{Introduction}
Small neutrino masses are observed in neutrino oscillation experiments. One of 
simple mechanisms to generate the small neutrino masses is the conventional 
Type-I seesaw mechanism~\cite{seesaw}, in which right-handed neutrinos $N_R^i$ 
($i$ denotes the generation) are introduced to the standard model (SM). The 
masses of light active neutrinos are described by $M_\nu\simeq-M_DM_R^{-1}M_D^T$ 
when $|(M_D)_{\alpha i}|\ll |(M_R)_{ij}|$, where $\alpha$ denotes the flavor, $M_D$ 
is the Dirac neutrino mass matrix given by the Yukawa coupling matrix of 
neutrinos $Y_\nu$ and the vacuum expectation value (VEV) of the Higgs $v$ 
($M_D\equiv Y_\nu v$), and $M_R$ is the Majorana mass matrix for $N_R$. When one 
takes the magnitude of the neutrino Yukawa couplings as $\mathcal{O}(1)$ like 
the top Yukawa coupling, a typical size of the right-handed neutrino Majorana 
mass becomes $M_R\sim\mathcal{O}(10^{14})$ GeV to generate the light active 
neutrino masses as $m_\nu \sim\mathcal{O}(0.1)$ eV. On the other hand, 
the right-handed neutrinos with the electroweak (EW) scale masses require relatively
 small $M_D$ of $\mathcal{O}(10^{-4})$ GeV to realize the active neutrino mass 
scale of $\mathcal{O}(0.1)$ eV in the Type-I seesaw mechanism. 

There are several extensions of the Type-I seesaw model. One extension is the 
inverse (double) seesaw mechanism~\cite{Wyler:1982dd,3,GonzalezGarcia:1988rw} 
with additional singlet fermions $S_\alpha$. In a basis of $(\nu_L^c, N_R, S)^T$ 
with three flavors (generations), a neutrino mass matrix is given by
\begin{eqnarray}
  \mathcal{M}=
   \left(
    \begin{array}{ccc}
     0       & M_D        & 0          \\
     M_D^T & 0           & M_{S}   \\
     0       & M_{S}^T &  \mu \\
    \end{array}
   \right).
     \label{original}
 \end{eqnarray}
When one assigns the lepton number one unit to $\nu_L, N_R$, and $S$ ($S$ has 
an opposite lepton number with respect to that of $N_R$), the Majorana mass 
terms of $S$ do not conserve the lepton number. Note that the absence of 
$(\mathcal{M})_{13}$ and $( \mathcal{M})_{31}$ in Eq.~(\ref{original}) are ensured
 by a field redefinition. Assuming $\mu \ll M_D < M_S$,  one can describe the 
neutrino mass as $M_\nu\simeq \mu M_D^2/M_S^2$ by diagonalizing the above mass 
matrix. For $M_D=10$ GeV and $M_S=$ 1 TeV, $\mu \simeq 1$ keV is required for 
generating the small active neutrino mass scale. 
In this model, one obtains the heavy Majorana neutrinos with masses as  $M_S\pm\mu /2$. Thus, 
when $\mu\ll M_S$, such neutrinos are degenerate in mass and can realize the 
resonant leptogenesis mechanism to generate the baryon asymmetry of the universe
 (BAU)~\cite{Gu:2010xc,Baldes:2013eva, Blanchet:2010kw, Basso:2012ti}. The 
explanations of neutrino experimental data and dark matter in the generic class 
of the inverse seesaw model have been discussed in~\cite{Abada:2014vea} and 
\cite{Abada:2014zra}, respectively.

In this work, we will discuss the inverse seesaw model realized by a 
``seesaw'' mechanism in the TeV scale physics. Our model has two 
kinds of new gauge singlet fermions $S_{1}$ and $S_{2}$ in addition to $N_R$, 
which corresponds to the $n=2$ multiple seesaw mechanism in 
Ref.~\cite{Xing:2009hx}. We will find that our model can naturally 
induce very small mass difference between heavy ($\sim$ TeV scale) neutrino 
states, which can also be responsible for a successful resonant leptogenesis.

\section{Inverse seesaw from ``seesaw"}

We discuss a realization of the inverse seesaw from the ``seesaw'' 
mechanism. The relevant Lagrangian is given by
 \begin{eqnarray}
  -\mathcal{L}=Y_\nu\tilde{H}\overline{L}N_R+Y_{S_1}\Phi_1\overline{N_R^c}S_{1}
                    +Y_{S_2}\Phi_2\overline{S_{1}^c}S_{2}
                    +\frac{M_\mu}{2}\overline{S_{2}^c}S_{2}+h.c.,
  \label{Lagrangian}
 \end{eqnarray}
where $\tilde{H}\equiv i\sigma_2H^\ast$, $H$ is the SM Higgs doublet, $L$ is the 
left-handed lepton doublet, $N_R$ is the right-handed neutrinos, $\Phi_1$ and 
$\Phi_2$ are gauge singlet scalars under the SM gauge groups, $S_{1}$ and $S_{2}$
 are gauge singlet fermions, and $M_\mu$ is a Majorana mass of $S_{2}$. 
Note that $S_{2}$ is added to the original inverse seesaw mechanism. Here 
details of additional symmetries in our model are not specified, 
but discussed later. In order to reproduce two (solar and atmospheric) mass 
scales of the active neutrinos, one must introduce at least two generations for 
the right-handed neutrino or the gauge singlet fermions. We omit the generation 
and flavor indices for fermions in Eq.~(\ref{Lagrangian}). After spontaneous 
gauge symmetry breaking, one can describe a neutrino mass matrix as
 \begin{eqnarray}
  \mathcal{M}=
   \left(
    \begin{array}{cccc}
     0    & M_D     & 0       & 0      \\
     M_D^T & 0      & M_{S_1}   & 0     \\
     0    & M_{S_1}^T & 0       & M_{S_2} \\
     0    & 0       & M_{S_2}^T & M_\mu
    \end{array}
   \right), 
 \label{mmISS}
 \end{eqnarray}
in the basis of $(\nu_L^c,N_R,S_{1},S_{2})^T$ where $M_D\equiv 
Y_\nu\langle H\rangle$, $M_{S_1}\equiv Y_{S_1}\langle \Phi_1\rangle$, 
$M_{S_2}\equiv Y_{S_2}\langle \Phi_2\rangle$, and these are described by 
matrices.\footnote{A similar structure of the mass matrix has been discussed in 
the three active and two sterile neutrinos model for 
the liquid scintillator neutrino detector anomaly~\cite{Mohapatra:2005wk}.} If one adds three generations for each singlet 
fermion, the neutrino mass matrix $\mathcal{M}$ is a $12\times12$ matrix.

When we assume that values of all matrix elements of $M_{S_2}$ are much smaller 
than those of $M_\mu$ ($(M_{S_2})_{jk}\ll(M_\mu)_{lm}$), we can diagonarize lower 
right $2\times2$ sub-matrix (integrate out $S_{2}$ field), i.e. utilize a 
``seesaw" mechanism. Then, the block diagonalization gives
 \begin{eqnarray}
  \mathcal{M}\rightarrow 
   \left(
    \begin{array}{cc}
     \bar{\mathcal{M}} & 0               \\
     0                      & \bar{M}_\mu
    \end{array}
   \right),
 \end{eqnarray}
with
 \begin{eqnarray}
  \bar{\mathcal{M}}\simeq
   \left(
    \begin{array}{ccc}
     0       & M_D        & 0                                \\
     M_D^T & 0            & M_{S_1}                            \\
     0       & M_{S_1}^T   & \mu
    \end{array}
   \right),~~~
  \bar{M}_\mu\simeq M_\mu,~~~
  \mu\simeq-M_{S_2}M_\mu^{-1}M_{S_2}^T,
  \label{l}
 \end{eqnarray}
in a basis. Notice that $ \bar{\mathcal{M}}$ takes the same form of the original
 inverse seesaw as in Eq.~(\ref{original}), and the smallness of $\mu$ can be 
naturally realized by the ``seesaw" mechanism, 
$\mu\simeq-M_{S_2}M_\mu^{-1}M_{S_2}^T$.

The mass matrix for the three active neutrinos $M_\nu$ can be obtained after the
 inverse seesaw as
 \begin{eqnarray}
  M_\nu\simeq -M_D(M_{S_1}^T)^{-1}M_{S_2}M_\mu^{-1}M_{S_2}^T(M_{S_1})^{-1}M_D^T,
 \end{eqnarray}
in the flavor basis of active neutrinos. The magnitude of the matrix elements of
 active neutrino is realized as
 \begin{eqnarray}
  M_\nu\sim\left(\frac{M_D}{10~{\rm GeV}}\right)^2
          \left(\frac{M_{S_1}}{1~{\rm TeV}}\right)^{-2}
          \left(\frac{M_{S_2}}{30~{\rm MeV}}\right)^2
          \left(\frac{M_\mu}{1~{\rm TeV}}\right)^{-1}0.1~{\rm eV}.
 \label{ISS}
 \end{eqnarray}
In the conventional (TeV scale) inverse seesaw mechanism, one should require 
$\mu$ in the mass matrix of Eq.~(\ref{original}) of $\mathcal{O}(1)$ keV scale. On the 
other hand, the realization of the inverse seesaw from ``seesaw"
Eq.~(\ref{mmISS}) needs $M_{S_2}$ 
of $\mathcal{O}(10)$ MeV scale in stead of keV scale. Therefore, the mass 
hierarchy among the singlet fermions in our model becomes small compared to the 
usual inverse seesaw model. Considering that light quarks and 
leptons have MeV-scale masses, the scale could be usable as a parameter of the 
model.

\section{Leptogenesis}

Next, we discuss a generation of the BAU. Our model includes 
several singlet Majorana fermions, and masses of some of them can be taken as 
$\mathcal{O}(1)$ TeV. Thus, the resonant leptogenesis~\cite{Pilaftsis:2003gt} 
might be possible in the model.

We start from the mass matrix Eq.~(\ref{l}). Since typical size of the matrix 
elements of $M_{S_1}$ is much larger than that of $\mu$, a mixing angle for a 
block diagonalization of lower right $2\times2$ sub-matrix of Eq.~(\ref{l}) is 
almost maximal. Thus, $\bar{\mathcal{M}}$ is rotated as
 \begin{eqnarray}
  \bar{\mathcal{M}}\rightarrow\bar{\mathcal{M}}'\simeq
   \left(
    \begin{array}{ccc}
     0             & M_D(1+\epsilon)/\sqrt{2} & M_D(1+\epsilon)/\sqrt{2} \\
     (M_D(1+\epsilon))^T/\sqrt{2} & M_{S_1}-\mu/2 & 0           \\
     (M_D(1+\epsilon))^T/\sqrt{2} & 0           & M_{S_1}+\mu/2 
    \end{array}
   \right),
 \label{8}
 \end{eqnarray}
up to order of $\mathcal{O}(\mu)$ in a basis of $(\nu_L^c,X_-,X_+)^T$ where 
$\epsilon\simeq\mu/(2\sqrt{2}M_{S_1})$, and eigenstates $X_\pm$ are described as 
$X_\pm\simeq(c_RN_R\mp c_{1R}S_{1}\mp c_1S_{2})/\sqrt{2}$ with $c_R\simeq 
c_{1R}\simeq1$ where $c_1$ is estimated as a typical ratio of matrix elements of 
$M_{S_2}$ and $M_\mu$, $c_1\simeq\mathcal{O}(M_{S_2}/M_\mu)$. The relevant 
Lagrangian for the resonant leptogenesis is given from Eq.~(\ref{Lagrangian}) as
 \begin{eqnarray}
  -\mathcal{L}\supset 
    Y_\nu^N\tilde{H}\overline{L}X_-
                      +Y_\nu^S\tilde{H}\overline{L}X_+  
                      +\frac{M_{S_1}-\mu/2}{2}\overline{X_-^c}X_-
                      +\frac{M_{S_1}+\mu/2}{2}\overline{X_+^c}X_++h.c.,
 \end{eqnarray}
for the eigenstates of $X_\pm$. Note that typical size of matrix elements of 
$Y_\nu^N$ and $Y_\nu^S$ is the same order as that of $Y_\nu$ in 
Eq.~(\ref{Lagrangian}), $Y_\nu^N\simeq Y_\nu^S\simeq(c_R/\sqrt{2})Y_\nu$ at the 
leading order. Hereafter we assume that 3$\times$3 matrices $M_{S_1}$ and $\mu$ 
are diagonal matrices, for simplicity. We also assume the hierarchical structure
 for $M_{S_1}$ as $m_{S_1}\equiv(M_{S_1})_{11}\ll(M_{S_1})_{22},(M_{S_1})_{33}$ so that 
the BAU can be induced by the decays of the first generation of $X_\pm$ 
($\equiv\chi_\pm$) whose masses are obtained as $m_{\chi_\pm}=m_{S_1}\pm 
\mu/2$.\footnote{Such a hierarchical mass structure among singlet Majorana 
fermions can be realized by several 
models~\cite{Kusenko:2010ik,Lindner:2010wr,Merle:2011yv,Takahashi:2013eva}.}

The lepton asymmetry from the decays of $\chi_-$ and $\chi_+$ is calculated 
as~\cite{Pilaftsis:1997jf, Gu:2010xc}
 \begin{eqnarray}
  \epsilon_\pm &=&
   \frac{\sum_\alpha\left[\Gamma(\chi_\pm\rightarrow L_\alpha+H^\ast)
                        -\Gamma(\chi_\pm\rightarrow L_\alpha^c+H)\right]}
        {\sum_\alpha\left[\Gamma(\chi_\pm\rightarrow L_\alpha+H^\ast)
                        +\Gamma(\chi_\pm\rightarrow L_\alpha^c+H)\right]} 
  \nonumber \\
  &\simeq& 
   \frac{{\rm  Im}(Y_\nu^{N \dag}Y_\nu^{S}Y_\nu^{N \dag}Y_\nu^{S})_{11}}{8\pi A_{\pm}}
   \frac{r}{r^2+\Gamma_\mp^2/m_{\chi_\mp}^2},
 \end{eqnarray}
where
 \begin{eqnarray}
  r = \frac{m_{\chi_+}^2-m_{\chi_-}^2}{m_{\chi_+} m_{\chi_-}}
        \simeq\frac{2\mu}{m_{S_1}},~~~
  A_+ = (Y_\nu^{S\dag}Y_\nu^S)_{11},~~~A_-=(Y_\nu^{N\dag}Y_\nu^N)_{11}, \label{r}
 \end{eqnarray}
and $\Gamma_{\pm}=A_{\pm} m_{\chi_\pm}/(8\pi)$ is the decay width of $\chi_\pm$. The 
baryon asymmetry is given by the lepton asymmetry as    
 \begin{eqnarray}
  \eta_B = -\frac{28}{79}\frac{0.3 \epsilon_\pm}{g_{*} K_\pm (\ln K_\pm)^{0.6}}, 
  \end{eqnarray}
where $g_{*}=106.75$ is the relative degree of freedom and 
$K_\pm=\Gamma_\pm/(2 H(T))|_{T=m_{\chi_\pm}}$ with the Hubble constant 
$H(T)=1.66 \sqrt{g_\ast}T^2/m_{\rm Pl}$. Note that the baryon asymmetry is enhanced
 for $(m_{\chi_+}-m_{\chi_-})\sim\Gamma_\pm /2$.

In order to obtain the baryon asymmetry by the decays of $\chi_\pm$, the 
$\chi_\pm$ should be decoupled at $T\sim m_{\chi_\pm}$, which is 
realized for the Yukawa couplings $(Y_\nu^N)_{\alpha 1}$ and  $(Y_\nu^S)_{\alpha 1}$ 
being $<\mathcal{O}(10^{-6})$. Under these conditions, the appropriate order of 
$r$ in Eq.~(\ref{r}) for the resonant leptogenesis is $r\sim 10^{-9}$, which can 
also be naturally realized in our model. Regarding with masses of additional 
scalars $\Phi_{1,2}$, these must be larger than the masses of $\chi_\pm$ of 
$\mathcal{O}(1)$ TeV. If those masses are smaller than the TeV scale, $\chi_\pm$
 decay into the scalars. As a result, the lepton asymmetry cannot be produced. 
On the other hand, the VEV of $\Phi_2$ should be larger than $\mathcal{O}(10)$ 
MeV to realize the inverse seesaw when $Y_{S_2}\leq\mathcal{O}(1)$. Such a 
hierarchy between the mass and VEV can be realized in the neutrino-philic Higgs 
model~\cite{Ma:2000cc} (see also~\cite{Haba:2010zi,Haba:2014taa}). The relevant 
scalar potential for the realization is, for example,
\begin{eqnarray}
 V &\supset& 
 -m_{\Phi_1}^2|\Phi_1|^2+m_{\Phi_2}^2|\Phi_2|^2
             -m^2(\Phi_1^\ast\Phi_2+\Phi_2^\ast\Phi_1)
             +\frac{\lambda_{\Phi_1}}{2}|\Phi_1|^4
             +\frac{\lambda_{\Phi_2}}{2}|\Phi_2|^4
             \nonumber \\
 && +\lambda_3|\Phi_1|^2|\Phi_2|^2
    +\lambda_4|\Phi_1\Phi_2|^2
    +\frac{\lambda_5}{2}
     \left[(\Phi_1^\ast\Phi_2)^2+(\Phi_2^\ast\Phi_1)^2
     \right], \label{V}
\end{eqnarray}
where $m_{\Phi_{1,2}}$, $m$, $\lambda_{\Phi_{1,2}}$, and $\lambda_{3,4,5}$ 
are all assumed to be real and positive, for simplicity. The stationary 
conditions $\partial V/\partial\langle\Phi_1\rangle=0$ and $\partial 
V/\partial\langle\Phi_2\rangle=0$ lead $|\langle\Phi_1\rangle|\simeq 
m_{\Phi_1}/\sqrt{\lambda_{\Phi_1}}$ and $|\langle\Phi_2\rangle|\simeq 
m^2|\langle\Phi_1\rangle|/m_{\Phi_2}^2$, respectively, where we assume 
$\lambda_{\Phi_1},\lambda_{3,4,5}\ll\mathcal{O}(1)$ and 
$\lambda_{\Phi_2}|\langle\Phi_2\rangle|^2\ll m_{\Phi_2}^2$. In addition, when one 
introduces the symmetry, which forbids the term 
$m^2(\Phi_1^\dagger\Phi_2+\Phi_1\Phi_2^\dagger)$ in Eq.~(\ref{V}), the hierarchy 
$m\ll m_{\Phi_{1,2}}$ seems to be natural. As a result, 
$|\langle\Phi_2\rangle|\ll|\langle\Phi_1\rangle|$ can be realized, where an 
assumption $\lambda_{\Phi_2}|\langle\Phi_2\rangle|^2\ll m_{\Phi_2}^2$ is consistent 
with this realization. Thus, one can have the hierarchy between the VEVs, 
$|\langle\Phi_1\rangle|\simeq 
m_{\Phi_1}/\sqrt{\lambda_{\Phi_1}}\gtrsim\mathcal{O}(1)$ TeV and 
$\langle\Phi_2\rangle=\mathcal{O}(10)$ MeV, when one takes $m=\mathcal{O}(10)$ 
GeV and masses of $\Phi_{1,2}$ as $\mathcal{O}(1)$ TeV. Here we take masses of 
$\Phi_{1,2}$ are larger than the masses of $\chi_\pm$. The above calculation is 
valid in this case. In our model, the other singlet fermions ($\simeq S_2$) can 
also decay into $L$ and $H$ through the mixing between $S_1$ and $S_2$. But the 
process cannot generate the sufficient magnitude of lepton asymmetry because 
$S_2$ is not degenerate with $N_R$ and $S_1$ state. Note that one does not need 
a fine tuning among masses of Majorana fermions to realize the BAU as seen 
below.

Figure~{\ref{fig1}} shows the baryon asymmetry as a function of $\mu$.
We assume the hierarchical structures for Yukawa couplings $Y_\nu^N$ and $Y_\nu^S$
so as to realize the out of thermal equilibrium of $\chi_\pm$ at the 
$T\sim m_{\chi_\pm}$: $|(Y_\nu^N)_{\alpha 1}|$, 
$|(Y_\nu^S)_{\beta 1}|<\mathcal{O}(10^{-6})$. The observed baryon asymmetry 
$\eta_B=6\times 10^{-10}$ is also shown by the horizontal line in 
Fig.~{\ref{fig1}}. In the calculation, we take 
$(Y_\nu^N)_{\alpha 1}=(1.0+0.1\,i)\times10^{-6}$, 
$(Y_\nu^S)_{\alpha 1}=(1.0+0.3\,i)\times10^{-6}$, and $M_{S_1}=1$ TeV as reference 
values. It is seen that the observed baryon asymmetry can be realized at 
$\mu\simeq1$ keV.
\begin{figure}
\hspace{4.3cm}
\begin{center}
\includegraphics[scale=1.]{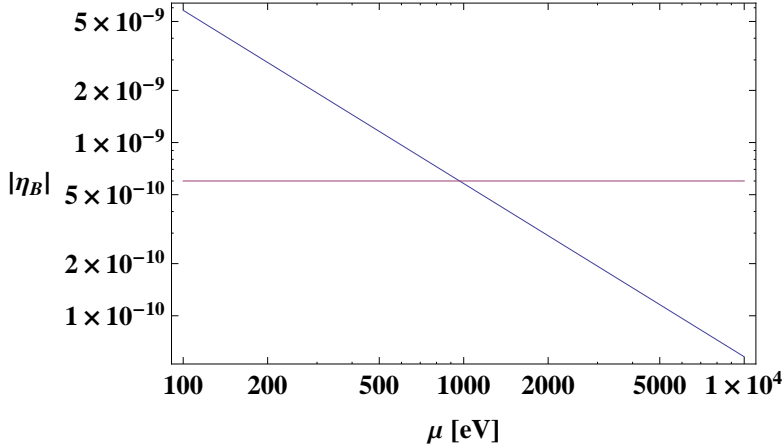}
\end{center}
\caption{Baryon asymmetry as a function of $\mu$.}
\label{fig1}
\end{figure}
The case of $\mu\simeq\mathcal{O}(1)$ keV is consistent with the realization of 
the small active neutrino mass in our model, i.e. 
$\mu\simeq M_{S_2}M_\mu^{-1}M_{S_2}^T\simeq\mathcal{O}(1)$ keV.

As discussed above, the favored scale of $\mu$ for the active neutrino mass can 
be realized by the ``seesaw" between $S_{1,2}$ fermions. In addition, the first 
generations of $S_1$ and $N_R$ (those mass eigenstates are $\chi_\pm$) play a 
role for generating the BAU via the resonant leptogenesis. In the case, the 
required size of mass degeneracy between $\chi_\pm$ for the resonant 
leptogenesis, $\mu\simeq\mathcal{O}(1)$ keV, can also be realized by the 
``seesaw" between $S_{1,2}$. The both realizations are non-trivial results in 
our model. 

\section{Signatures for LHC experiment}

We discuss signatures of this model at the LHC experiment. This model 
can induce lepton number violating processes. One interesting process is the 
like-sign dilepton production, $q\bar q'\rightarrow \ell^\pm\ell^\pm W^\mp$, where
 the lepton number conservation is violated by two units, $\Delta L=2$, due to 
the Majorana nature of neutrinos. 
References~\cite{Han:2006ip,delAguila:2007em,delAguila:2008cj,Atre:2009rg,Chen:2011hc,Helo:2013esa,Das:2014jxa,Izaguirre:2015pga} 
explore this process at the LHC experiment in the SM with the right-handed 
Majorana neutrinos (see also~\cite{Deppisch:2015qwa} for a review of the 
collider phenomenology with the right-handed and sterile Majorana 
neutrinos).\footnote{The analysis of this process is given in 
Ref.~\cite{Das:2012ze} for the inverse seesaw model in the context of the 
next-to-minimal supersymmetric SM.} According to Ref.~\cite{Atre:2009rg}, it is 
found that there is $2\sigma~(5\sigma)$ sensitivity for the $\mu^\pm\mu^\pm$ 
modes in the mass range of a Majorana neutrino of $10~\mbox{GeV}\leq 
m_\chi\leq350~(250)~\mbox{GeV}$ at the 14 TeV LHC experiment with 100 fb$^{-1}$.
Regarding with the inverse seesaw case, the singlet neutrinos and fermions are 
pseudo-Dirac neutrinos due to a small Majorana mass $\mu$, and the neutrinos 
contain tiny Majorana state. The ratio of the Majorana state is typically 
determined by 
$\mu/M_{S_1}\simeq1~\mbox{keV}/1~\mbox{TeV}\simeq\mathcal{O}(10^{-9})$. Thus, 
since the like-sign dilepton production process in the inverse seesaw case is 
suppressed by $(\mu/M_{S_1})^2\simeq\mathcal{O}(10^{-18})$ compared with the 
results of 
Refs.~\cite{Han:2006ip,delAguila:2007em,delAguila:2008cj,Atre:2009rg}, the 
signatures of the process in the inverse seesaw case cannot reach at the 
sensitivity at the LHC experiment.

Similarly, for the other singlet fermions ($\simeq S_2$) with 
the lepton number violating Majorana mass of $\mathcal{O}(1)$ TeV in our model,
the result of analysis in Refs.~\cite{Han:2006ip,delAguila:2007em, delAguila:2008cj, Atre:2009rg} 
cannot simply be adopted. Since the like-sign dilepton production process is 
induced through  
the mixing between $S_1$ and $S_2$ in addition to the mixing of 
the pseudo-Dirac states of $N_R$ and $S_1$ mentioned above,
the amplitude is typically suppressed by 
$(M_{S_2}/M_\mu)^2(\mu/M_{S_1})^2\simeq(10~\mbox{MeV}/1~\mbox{TeV})^2(1~\mbox{keV}/1~\mbox{TeV})^2\simeq\mathcal{O}(10^{-28})$. 
Therefore, the collider signatures of the these singlet fermions in our model
cannot also reach at the sensitivity at the LHC experiment.

The above discussion can be generalized to the multiple seesaw 
models~\cite{Xing:2009hx}. For the $n=2k+1$ ($k=0, 1, 2,\cdots$) multiple seesaw
 models ($k=0$ is the conventional inverse seesaw model), the active neutrino 
mass matrix in the $n=2k+1$ ($k\geq1$) multiple seesaw models is given by
 \begin{eqnarray}
  M_\nu=M_D\left[\prod_{i=1}^k(M_{S_{2i-1}}^T)^{-1}M_{S_{2i}}\right]
       (M_{S_{2k+1}}^T)^{-1}M_\mu(M_{S_{2k+1}})^{-1}\left[\prod_{i=1}^k
       (M_{S_{2i-1}}^T)^{-1}M_{S_{2i}}\right]^TM_D^T,
 \end{eqnarray}
where $n$ denotes the number of gauge singlet fermions $S$ without the number of
 generation (flavor) and $M_\mu$ is the lower right element of the 
$(n+2)\times(n+2)$ generalized neutrino mass matrix. The like-sign dilepton 
production process is suppressed by $(M_\mu/M_{S_n})^2$ in all models of $n=2k+1$ 
multiple seesaw with $M_\mu\simeq\mathcal{O}(1)~\mbox{keV}\ll 
M_{S_1}\simeq\cdots\simeq M_{S_{n-1}}$. On the other hand, for the $n=2k$ 
($k=1, 2,\cdots$) multiple seesaw models ($k=1$ case is our model), the active 
neutrino mass matrix can be given by
 \begin{eqnarray}
  M_\nu=-M_D\left[\prod_{i=1}^k(M_{S_{2i-1}}^T)^{-1}M_{S_{2i}}\right]M_\mu^{-1}
           \left[\prod_{i=1}^k(M_{S_{2i-1}}^T)^{-1}M_{S_{2i}}\right]^TM_D^T.
 \end{eqnarray}
The amplitude of the like-sign dilepton production process is suppressed 
by $(M_{S_n}/M_\mu)^2\times(1~\mbox{keV}/M_{S_1})^2$ in all models of $n=2k$ 
multiple seesaw with $M_{S_n}\ll M_\mu\simeq M_{S_1}\simeq\cdots\simeq M_{S_{n-1}}$. 
Note that since the $n=2k$ multiple seesaw model is reduced to the inverse 
seesaw model, there is an additional suppression $(M_{S_n}/M_\mu)^2$ in the $n=2k$
 cases compared with the $n=2k+1$ multiple seesaw models.\footnote{In 
Refs.~\cite{BhupalDev:2012zg,Bandyopadhyay:2012px,Arganda:2014dta}, the authors 
have discussed the Higgs signatures via the large Yukawa couplings in the 
inverse seesaw model at the LHC. The second and third generations of $X_\pm$ in 
our model might be adopted to the discussion although the Yukawa couplings of 
the first generations are too small to lead the sufficient magnitude of the 
signals.}

\section{Summary}

We have discussed the inverse seesaw model realized by a ``seesaw'' mechanism. 
The conventional inverse seesaw model requires the lepton number violating 
Majorana mass of $\mu\simeq\mathcal{O}(1)$ keV to achieve the light active 
neutrino mass scale when the Dirac masses are taken as $M_D=10$ GeV and $M_S=1$ 
TeV (see Eq.~\eqref{original}). The hierarchy among mass scales in the 
conventional inverse seesaw model is given by $M_S/\mu\simeq\mathcal{O}(10^9)$. 
On the other hand, in our model the Majorana mass is $M_\mu\simeq\mathcal{O}(1)$
 TeV for the Dirac masses of $M_D=10$ GeV, $M_{S_1}=1$ TeV, and $M_{S_2}=30$ MeV 
(see Eqs.~\eqref{mmISS} and \eqref{ISS}). Thus, the magnitude of mass hierarchy 
in the model can be decreased to $M_\mu/M_{S_2}\simeq\mathcal{O}(10^5)$, which is 
due to the ``seesaw'' mechanism between $S_1$ and $S_2$ singlet fermions.

We have also considered a leptogenesis scenario with a mass degeneracy for 
generating the BAU, the so-called resonant leptogenesis. The scenario can be 
realized by the keV scale mass degeneracy between the first generations of the 
right-handed neutrino and one of the singlet fermions. We have shown that such 
mass degeneracy can also be realized by the ``seesaw'' in our model, and thus 
the successful resonant leptogenesis is achieved. Regarding the signatures of 
$q\bar q'\rightarrow \ell^\pm\ell^\pm W^\mp$ processes at the LHC experiment, our 
model cannot reach at the sensitivity at the LHC due to the significant 
suppression by the mixings between the singlet fermions.
 
Finally, we comment on a realization of our model. One simple way to obtain our 
model is to introduce a symmetry. In Ref.~\cite{Xing:2009hx}, the global 
$U(1)\times \mathbb{Z}_{2N}$ symmetry for realizing the multiple seesaw models 
have been discussed. Following that, our model (the $n=2$ multiple seesaw model)
 can be obtained by imposing the global $U(1)\times \mathbb{Z}_{6}$ symmetry. 
Here the global $U(1)$ symmetry is identified with the lepton number, $U(1)_L$, 
and a charge assignment under the symmetry is given in Tab.~\ref{tab1}. Note 
that the Majorana mass term $(M_\mu/2)\overline{S_2^c}S_2$ induces the lepton 
number violation.
\begin{table}
\begin{center}
\begin{tabular}{c|c|c|c|c|c|c|c}
\hline
& $L$ & $H$ & $N_R$ & $S_{1R}$ & $S_{2R}$ & $\Phi_1$ & $\Phi_2$ \\
\hline\hline
$U(1)_L$ & $+1$ & $0$ & $+1$ &$-1$ &$+1$ &$0$ &$0$ \\
\hline
$\mathbb{Z}_6$ & $\omega_6$ & $+1$ & $\omega_6$ &$\omega_6^2$ &$\omega_6^3$ &$\omega_6^3$ &$\omega_6$ \\
\hline
\end{tabular}
\end{center}
\caption{Charge assignment~\cite{Xing:2009hx} for our model where 
$\omega_6\equiv e^{i\pi/3}$.}
\label{tab1}
\end{table}

\subsection*{Acknowledgement}

The authors thank H. Ishida and Y. Yamaguchi for helpful discussions, and thank 
O. Seto for giving important comments on the discussion of leptogenesis. The 
work of N.~H. is supported in part by the Grant-in-Aid for Scientific Research 
(Grants  No. 24540272, No. 15H01037, and No. 26247038.) and the work of M.~A. is
 supported in part by the Grant-in-Aid for Scientific Research (Grants No. 
25400250 and No. 26105509).

\end{document}